\documentclass[11pt]{article}

\usepackage[utf8]{inputenc}
\usepackage[T1]{fontenc}
\usepackage{arxiv}

\usepackage{amsmath}
\usepackage{amssymb}
\usepackage{amsthm}
\allowdisplaybreaks
\usepackage{graphicx}
\usepackage{booktabs}
\usepackage{float}
\usepackage{caption}
\usepackage{setspace}
\usepackage{indentfirst}
\usepackage[authordate,backend=biber,natbib=true]{biblatex-chicago}
\usepackage[hidelinks]{hyperref}

\DeclareMathOperator{\Cov}{cov}
\DeclareMathOperator{\Var}{var}

\DeclareMathOperator{\Sd}{sd}
\newcommand{\bu}[1]{\mathbf{#1}}
\newcommand{\bias}{\widehat{\text{bias}}}

\makeatletter
\renewcommand{\@seccntformat}[1]{\csname the#1\endcsname.\hspace{0.5em}}
\makeatother

\title{Omitted variable bias sensitivity analysis with clustered treatment
assignment}

\author{Anton Strezhnev\thanks{First draft: July 13, 2026. Support for this research was
provided by the Office of the Vice Chancellor for Research at the University of
Wisconsin-Madison with funding from the Wisconsin Alumni Research Foundation
(WARF). Claude Opus 4.8 was used to assist with developing the software
implementation, writing the replication code, reviewing the math and porting from Typst to .tex. All
errors are my own. A fork of the \texttt{sensemakr} R package implementing the
cluster adjustment methods in this paper is available at
\href{https://github.com/astrezhnev/sensemakr}{github.com/astrezhnev/sensemakr}.}\\
\small University of Wisconsin-Madison, Department of Political Science.
\texttt{strezhnev@wisc.edu}}

\date{July 13, 2026}

\begin{document}

\maketitle

\begin{abstract}
\citet{cinelli2020making} develops a sensitivity analysis method for the linear
regression model that parameterizes omitted variable bias in terms of two
partial $R^2$ parameters capturing the residual variation explained by an
omitted confounder in the treatment and outcome respectively. This method is
often applied to regressions fit to unit-level data when treatment is assigned
at a higher level of aggregation - as in clustered observational designs. This
paper shows that despite the numerical equivalence of the unit-level regression
and an appropriately weighted cluster-aggregated regression for estimating the
treatment effect, the sensitivity analysis procedure yields different
conclusions depending on the chosen level of analysis. The outcome-confounder
partial $R^2$ reflects both between- and within- group variation but the latter
is irrelevant to omitted variable bias as it by construction cannot be explained
by a group-level confounder. Straightforward corrections to the robustness value
and the extreme scenario analysis from the unit-level regression using Pearson's
partial-$\eta$ recover equivalence between these two approaches. The paper
concludes with a point of caution when benchmarking against unit-level
covariates and recommends always including cluster-level averages of these
covariates as regressors \citep{mundlak1978pooling}.
\end{abstract}


\section{Introduction}
\label{sec:intro}

One of the central challenges facing researchers when estimating causal effects
using selection-on-observables designs is defending the core identifying
assumption that there are no unobserved confounders of treatment and outcome.
Typically such designs will attempt to control for as many observed confounders
as possible, but unlike randomized experiments there is no guarantee that
treated and control groups are balanced on unadjusted-for correlates of the
outcome. Given an estimated effect, it is common to ask how ``strong'' of an
unobserved confounder would explain away the result. Sensitivity analysis
methods provide a structured way for researchers to quantify this by introducing
a parameter or set of parameters that define the degree of confounding present.
Such approaches date back to \citet{cornfield1959smoking} which asked how
strongly an unobserved trait would have to be associated with both smoking and
lung cancer to account for their observed association.

This paper focuses on sensitivity analysis for the linear regression model, a
ubiquitous method in applied research. Researchers regress an outcome on the
treatment of interest and some set of observed covariates and wish to interpret
the coefficient on treatment as a causal effect. In this setting, the familiar
omitted variables bias formula characterizes how the estimated effect would
change if a relevant confounder happened to be excluded from the regression. It
is common to see researchers informally ``sign the bias'' by reasoning about the
directions of the two components of omitted variables bias -- the
treatment-confounder and outcome-confounder associations. Formal sensitivity
analysis methods within the linear regression framework provide a means of
quantifying how large the confounding would have to be to substantially alter
the result \citep{frank2000impact, frank2013would, hosman2010sensitivity}.
Alternatively, a number of papers have proposed methods of \emph{benchmarking}
against some observed baseline \citep{altonji2005evaluation, oster2019unobservable}.
However, until recently, such sensitivity analyses had seen little adoption among
applied researchers due to the difficulties in selecting sensitivity parameters
that could be reasoned about substantively, the challenges of informal
benchmarking, the absence of standardized reporting statistics that could be
compared across studies and existing methods' reliance on strong parametric
assumptions.

\citet{cinelli2020making} provides the first formal treatment of the linear
regression omitted variables bias that requires no parametric assumptions on the
unobserved confounder and allows researchers to reason about the confounder in
terms of scale-free and bounded partial-$R^2$ parameters -- a parameterization
first introduced by \citet{imbens2003sensitivity} -- which capture the share of
residual variation in treatment and in outcome explained by a hypothetical
confounder. It develops a set of formal bounds for benchmarking the strength of a
hypothetical confounder against the variation explained by the observed
covariates, and gives a standard set of simple to report robustness statistics
that can accompany any regression table. One such statistic is the
\emph{robustness value} which reports the minimum strength that a confounder
equally associated with treatment and outcome would need in order to reduce the
estimate to zero. Implemented in the easy to use \texttt{sensemakr} software
\citep{cinelli2019sensemakr}, this method has seen significantly wider adoption
than many of the earlier methods, particularly in the social sciences where
linear regression remains a standard tool for covariate adjustment.

Other sensitivity analysis methods outside of the OLS tradition use alternative
parameterizations for other estimation strategies. Design-based bounds for the
\emph{matched pair} design \citep{rosenbaum1987sensitivity,
rosenbaum2002observational, rosenbaum2005heterogeneity} consider deviations from
a hypothetical ideal randomized experiment. Sensitivity to unobserved confounding
is parameterized by bounding the ratio $\Gamma$ by which the odds of treatment
assignment may differ between two units with identical observed covariates. The
analysis reports the range of randomization-inference $p$-values consistent with
that bound. A second family addresses the sensitivity of \emph{weighting
estimators} by assessing the extent to which the estimated weights diverge from
the ``ideal'' weights necessary to eliminate confounding. These approaches
include the marginal sensitivity model \citep{tan2006distributional,
zhao2019sensitivity} and the variance-based sensitivity model
\citep{huang2025variance}. A third family parameterizes the bias directly through
the \emph{confounding function}, the difference in common potential outcomes
among units with different observed treatment assignments
\citep{robins1999association, robins2000sensitivity, brumback2004sensitivity,
blackwell2014selection}. The sensitivity analysis is conducted by adjusting the
outcome by subtracting this function, evaluated at varying levels of the
sensitivity parameter, and re-estimating. A fourth approach directly
\emph{parameterizes} the latent confounder and its joint relationship with
treatment and outcome \citep{rosenbaum1983assessing, imbens2003sensitivity,
carnegie2016assessing, zheng2025copula}. Conversely, a fully model-agnostic
approach is the E-value \citep{ding2016sensitivity, vanderweele2017sensitivity},
the modern descendant of \citet{cornfield1959smoking} and the analogue of the
\citet{cinelli2020making} \emph{robustness value} for the risk ratio scale.

The primary contribution of this paper is to study the application of the
\citet{cinelli2020making} approach to the clustered observational design where
units are grouped into some higher level of aggregation (e.g. school, county,
village) at which treatment is assigned. Units that share a cluster share the
same treatment. In this setting, it is not uncommon for applied researchers to
estimate treatment effects with regressions fit to unit-level data. For example,
\citet{reny2026rising}, re-analyzed in Section~\ref{sec:replication}, estimates
the effect of a county-level measure of exposure to rising sea levels on an
individual-level measure of support for climate mitigation measures using a
regression fit to the individual survey responses and incorporating individual
level covariates such as partisanship. Although most of these studies now
correctly address the variance estimation issues that clustering creates by
using an appropriate estimator for the standard errors
\citep{liang1986longitudinal, abadie2023should}, there is less clarity regarding
how one should approach the sensitivity analysis for such a regression. Many
studies simply apply the \citet{cinelli2020making} method directly to the
unit-level regression, report robustness values, and benchmark confounding
against other covariates from that regression.

I show that such commonly-reported summaries are misleading and can actually
\emph{understate} the robustness of the design. This is because the unit-level
regression involves two partial-$R^2$ sensitivity parameters measured on
asymmetric scales. In a clustered observational design, confounding can only come
from between-cluster variation. Therefore, while a worst-case confounder can
explain 100\% of treatment variation, it cannot explain 100\% of outcome
variation. The upper bound on the unit-level partial-$R^2$ between outcome and
confounder is the ratio of between-cluster outcome variation to the total
variation. Because of this unaccounted-for compression in the range of the
outcome-confounder sensitivity parameter, the \citet{cinelli2020making}
\emph{robustness value}, which sets the two partial-$R^2$ sensitivity parameters
equal to one another, does not truly reflect ``equal confounding.'' At every given
partial-$R^2$, the outcome-confounder strength is \emph{greater} than the
treatment-confounder strength except when there is absolutely no within-cluster
variation in the outcome. Likewise, the \emph{extreme scenario} analysis,
considers the impossible case where a confounder explains \emph{all} residual
outcome variation rather than just the remaining variation across clusters.

Two earlier papers study the same setting of cluster-level treatment assignment,
but come from different sensitivity analysis traditions.
\citet{hansen2014clustered} considers a Rosenbaum-style sensitivity analysis and
reaches a conclusion similar to that of this paper. It shows that clustering in
treatment systematically \emph{reduces} sensitivity to unmeasured confounding,
because such a confounder is constrained by the structure of treatment assignment
in seeking out the most favorable individuals for treatment. I obtain an
analogous result for the omitted variables bias sensitivity analysis. Clustered
treatment assignment limits the variation that the confounder can explain and
improves ``robustness'' much in the same way that residualizing the outcome on a
purely prognostic covariate would. A more recent paper,
\citet{huang2026sensitivity} works in the weighting tradition and is the closest
paper to this one in that it analyzes the clustered observational design and
extends the variance-based sensitivity model to weighting-based adjustments,
distinguishing between unit-level and cluster-level covariates in the
construction of the weights.

I further show that the entire discrepancy between the unit-level and
cluster-level sensitivity analyses is governed by a single quantity $\eta^2$ that
denotes the share of the regression's residual outcome variance that can be
explained by the cluster indicators. This quantity originates in the statistics
literature as Pearson's ``correlation ratio'' \citep{pearson1905general} and has
a deep history in ANOVA \citep{kennedy1970eta, cohen1973eta} as well as
non-parametric regression \citep{doksum1995nonparametric}. Correcting the
proposed ``routine reporting'' measures obtained from unit-level regression using
this scaling factor yields exact equivalence between the unit-level and
cluster-level sensitivity analyses and illustrates that the uncorrected
unit-level sensitivity analyses are overly conservative. I then show that the
proposed benchmarking bounds in \citep{cinelli2020making} are equivalent when
applied to the unit-level and cluster-level regressions as long as the covariate
used for benchmarking is measured at the cluster level. However, this equivalence
breaks down when benchmarking against a unit-level covariate that varies both
within and between clusters. A simple fix to conventional practice -- including
group-level averages of individual-level covariates as an additional regressor
\citep{mundlak1978pooling} -- makes the within-cluster variation irrelevant and
facilitates valid benchmarking against a cluster-level confounder.

I apply the proposed correction to a sensitivity analysis conducted in
\citet{reny2026rising}, which estimates the effect of county-level sea rise
exposure on individual attitudes towards climate mitigation. The reported
robustness value of $7.4\%$ for the main analysis is very conservative since only
about a third of the residual outcome variation is between counties
($\eta^2 = 0.32$). Correcting for this raises the robustness value from $7.4\%$
to $12.7\%$ and the extreme scenario threshold from $0.59\%$ to $1.83\%$, lifting
the latter far above the treatment-outcome benchmark covariate: party
identification. Benchmarking against the raw individual-level covariate is, in
this case, conservative relative to benchmarking against the cluster means due to
the high degree of within-county correlation between party ID and climate
attitudes.

The remainder of the paper proceeds as follows. Section~\ref{sec:setup} sets up
the clustered observational design and reviews the \citet{cinelli2020making}
framework as applied to the cluster-level regression. Section~\ref{sec:ovb}
derives the correction factor $\eta^2$ for the unit-level regression and the
corresponding adjustments to the robustness value and the extreme scenario
analysis. Section~\ref{sec:individualcovariates} turns to benchmarking,
establishing the equivalence of benchmarks against group-level covariates under
either regression and characterizing the distortions induced by benchmarking
against individual-level covariates included in the regression without their
cluster-averages. Section~\ref{sec:replication} presents the reanalysis and
illustrates how published sensitivity analyses for regressions with clustered
treatment likely understate robustness. Section~\ref{sec:conclusion} concludes
with implications for related settings and recommendations for researcher
practice.

\section{Setup}
\label{sec:setup}

I start by reviewing the OVB sensitivity analysis setup introduced in
\citet{cinelli2020making} in the setting with a cluster-level treatment. When
applied to a cluster-level regression, the sensitivity analysis procedure remains
unchanged. Imagine that researchers are interested in comparing the coefficient
estimates from two regression specifications, a \emph{full} model that would
include a hypothetical confounder $Z$ and a \emph{restricted} specification that
omits it. The data consist of $N$ observations nested into $C$ clusters,
$N > C$. Index units by $i$ and clusters by $g$ with $g(i)$ denoting the cluster
that unit $i$ is a member of. Let $N_g$ denote the number of units in cluster
$g$. For simplicity, I will focus on the design with \emph{balanced} cluster
sizes, such that $C \times N_g = N$, but extension to an unbalanced design is
straightforward by weighting the cluster-level regression by $N_g$. An outcome
$Y_i$ is observed for each unit and treatment $D_g$ for each cluster.

I study the \emph{clustered observational design} characterized in
\citet{huang2026sensitivity}, focusing specifically on the ``cluster only
design'' in which units' cluster memberships are unaffected by treatment
assignment. Under this design, treatment-outcome confounding is driven by
cluster-level covariates (which can include aggregates of individual-level
covariates). Let $\bu{X}$ denote the $C$ by $P$ matrix of observed cluster-level
covariates (including the intercept) and $X_g$ denote the $P$-length vector of
observed cluster-level covariates for cluster $g$. $Z_g$ denotes a cluster-level
``unobserved'' confounder.

First consider the \emph{cluster-level} regression which regresses the cluster
averages $\bar{Y}_g = \frac{1}{N_g} \sum_{i:g(i) = g} Y_i$ on the treatment and
covariates. A researcher wishing to control for all sources of confounding would
estimate the \emph{full} ordinary least squares regression:

\begin{equation}\label{eq:clusterfull}
\bar{Y}_g = \hat{\tau} D_g + X_g^T \hat{\beta} + \hat{\gamma} Z_g
+ \hat{\epsilon}_g^{\text{full}}
\end{equation}

However, because $Z$ is unobserved, the researcher instead estimates the
\emph{restricted} regression:

\begin{equation}\label{eq:clusterres}
\bar{Y}_g = \hat{\tau}_{\text{res}} D_g + X_g^T \hat{\beta}_{\text{res}}
+ \hat{\epsilon}_g^{\text{res}}
\end{equation}

\citet{cinelli2020making} characterize the discrepancy between the unrestricted
and restricted coefficient estimates on the treatment
$\bias = \hat{\tau} - \hat{\tau}_{\text{res}}$. Applying the classic
Frisch-Waugh-Lovell theorem \citep{frisch1933partial, lovell1963seasonal}, they
derive the classic omitted variables bias formula in terms of sample covariances
and variances. Following the original notation, let $\perp$ denote the residual
of the variable from its linear projection onto the variables denoted in the
superscript.

\begin{equation}\label{eq:ovbderiv}
\begin{aligned}
\hat{\tau}_{\text{res}}
&= \frac{\Cov\left(D_g^{\perp \bu{X}}, \bar{Y}_g^{\perp \bu{X}}\right)}
{\Var\left(D_g^{\perp \bu{X}}\right)} \\
&= \frac{\Cov\left(D_g^{\perp \bu{X}},
\hat{\tau} D_g^{\perp \bu{X}} + \hat{\gamma} Z_g^{\perp \bu{X}}\right)}
{\Var\left(D_g^{\perp \bu{X}}\right)} \\
&= \hat{\tau} + \hat{\gamma}
\frac{\Cov\left(D_g^{\perp \bu{X}}, Z_g^{\perp \bu{X}}\right)}
{\Var\left(D_g^{\perp \bu{X}}\right)} \\
&= \hat{\tau} + \hat{\gamma}\hat{\theta}
\end{aligned}
\end{equation}

where $\hat{\theta} \equiv
\frac{\Cov\left(D_g^{\perp \bu{X}}, Z_g^{\perp \bu{X}}\right)}
{\Var\left(D_g^{\perp \bu{X}}\right)}$ is the coefficient on treatment from a
cluster-level regression of the unobserved confounder on treatment and the
observed covariates. The OVB formula gives us $\bias = \hat{\gamma}\hat{\theta}$,
the product of the association between the unobserved confounder and the outcome
net of the observed covariates $\hat{\gamma}$ and the association between
treatment and the unobserved confounder $\hat{\theta}$. While it is possible to
implement a sensitivity analysis in terms of these two parameters directly, the
major innovation of \citet{cinelli2020making} is to re-parameterize the OVB into
a \emph{scale-invariant} form using partial $R^2$s. Denote the partial $R^2$ of
the regression of $Z_g$ on $D_g$ adjusting for $\bu{X}$ as
$R^2_{Z_g \sim D_g \mid \bu{X}}$. The absolute omitted variables bias can be
written as:

\begin{equation}\label{eq:biascluster}
\left|\bias\right| = \sqrt{ R^2_{\bar{Y}_g \sim Z_g \mid D, \bu{X}}
\times \frac{R^2_{D_g \sim Z_g \mid \bu{X}}}{1 - R^2_{D_g \sim Z_g \mid \bu{X}}} }
\times \frac{\Sd\left(\bar{Y}_g^{\perp D, \bu{X}}\right)}
{\Sd\left(D_g^{\perp \bu{X}}\right)}
\end{equation}

The re-parameterization yields two sensitivity analysis parameters
$R^2_{\bar{Y}_g \sim Z_g \mid D, \bu{X}}$ and $R^2_{D_g \sim Z_g \mid \bu{X}}$
that are bounded between $0$ and $1$. The first reflects the proportion of the
variance explained in the cluster-averaged outcome by a hypothetical unobserved
confounder, conditional on treatment and observed covariates. The second reflects
the proportion that confounder explains in \emph{treatment} conditional on
observed covariates. \citet{cinelli2020making} characterize this as a scale
invariant ``bias factor''

\begin{equation}
\text{BF} \equiv \left| R_{\bar{Y}_g \sim Z_g \mid D, \bu{X}}
\times \frac{R_{D_g \sim Z_g \mid \bu{X}}}
{\sqrt{1 - R^2_{D_g \sim Z_g \mid \bu{X}}}} \right|
= \left| R_{\bar{Y}_g \sim Z_g \mid D, \bu{X}}
\times f_{D_g \sim Z_g \mid \bu{X}} \right|
\end{equation}

where $f = \sqrt{\frac{R^2}{1-R^2}}$ denotes Cohen's $f$ \citep{cohen1988}. This
bias factor is multiplied by a scale correction
$\Sd(\bar{Y}_g^{\perp D_g, \bu{X}})/\Sd(D_g^{\perp \bu{X}})$,
the ratio of the residual standard deviations of outcome and treatment, to yield
the omitted variable bias on the level of the original outcome. Furthermore, the
relative size of the bias induced by confounding compared to the magnitude of the
estimate $\bias/\hat{\tau}_{\text{res}}$ can be written in terms of this bias
factor scaled by the partial-$f$ of the outcome on treatment and observed
covariates.

\begin{equation}\label{eq:relativebiascluster}
\text{relative bias} = \frac{\left| R_{\bar{Y}_g \sim Z_g \mid D, \bu{X}}
\times f_{D_g \sim Z_g \mid \bu{X}} \right|}
{\left| f_{\bar{Y}_g \sim D_g \mid \bu{X}} \right|}
\end{equation}

In addition to the popular contour plot, which summarizes the adjusted treatment
effect under each hypothetical combination of the two sensitivity parameters,
\citet{cinelli2020making} define two summary robustness statistics that are
recommended as a standard reporting diagnostic alongside regression estimates:
the \emph{robustness value} and the \emph{extreme scenario analysis}. The
robustness value $\text{RV}_q$ is defined as the strength of association that a
confounder equally associated with treatment and outcome
($\text{RV}_q = R^2_{\bar{Y}_g \sim Z_g \mid D, \bu{X}}
= R^2_{D_g \sim Z_g \mid \bu{X}}$) would need to reduce the estimated effect by a
factor of $100\% \times q$.\footnote{\texttt{sensemakr} defaults to \texttt{q=1}
in its standard reporting.} They show that this can be straightforwardly computed
using Cohen's $f$ of the outcome on treatment given observed covariates scaled by
the reduction factor $q$. Define
$f_q \equiv q \times \left| f_{\bar{Y}_g \sim D_g \mid \bu{X}} \right|$. The
robustness value is then equal to

\begin{equation}\label{eq:rvcluster}
\text{RV}_q = \frac{1}{2}\left(\sqrt{f_q^4 + 4 f_q^2} - f_q^2\right)
\end{equation}

This quantity is related to other single-number robustness quantities such as the
``E-value'' \citep{vanderweele2017sensitivity}, which capture the general notion
of a ``minimum strength'' confounder necessary to overturn the results. That is,
to achieve the necessary amount of bias, a confounder could not have a
\emph{lower} partial-$R^2$ than $\text{RV}_q$ with respect to the outcome without
compensating by having a \emph{higher} partial-$R^2$ than $\text{RV}_q$ with
respect to the treatment (and vice-versa).

The other proposed reporting diagnostic, the \emph{extreme scenario} analysis,
instead considers robustness to a ``worst case'' threat to identification: one
where the confounder is assumed to explain the entire residual variation in the
outcome: $R_{\bar{Y}_g \sim Z_g \mid D, \bu{X}} = 1$. In this case, it is
straightforward to obtain this quantity, denoted $\text{XRV}$ from
Equation~\ref{eq:relativebiascluster}. Setting
$R_{\bar{Y}_g \sim Z_g \mid D, \bu{X}}$ and solving for the value of
$R^2_{D_g \sim Z_g \mid \bu{X}}$ that would set the relative bias to $1$ yields
$\text{XRV} = R^2_{\bar{Y}_g \sim D_g \mid \bu{X}}$.

\section{Adjusting sensitivity parameters for clustered designs}
\label{sec:ovb}

I now consider the application of the OVB sensitivity analysis to the unit-level
regression in which variables are indexed by $i$. Note that any terms indexed
\emph{only} by $g$ refer to the cluster-level regression while those that include
$i$ correspond to the unit-level regression. I again evaluate the bias in terms
of the difference in coefficient estimates across a \emph{full} specification and
a \emph{restricted} specification. Denote coefficients estimated with the
unit-level regression via the superscript $\text{U}$. Suppose instead of
Equation~\ref{eq:clusterres}, the analyst fit the following restricted
regression:

\begin{equation}\label{eq:unitrestricted}
Y_i = \hat{\tau}^{\text{U}}_{\text{res}} D_{g(i)}
+ X_{g(i)}^T \hat{\beta}_{\text{res}}^{\text{U}}
+ \hat{\epsilon}_i^{\text{res}}
\end{equation}

Write the omitted variable bias for $\hat{\tau}_{\text{res}}^{\text{U}}$ as
before, applying Frisch-Waugh-Lovell to the unit level regression. Write the
outcome in terms of its \emph{between} cluster variation $\bar{Y}_{g(i)}$ and its
\emph{within} cluster residual $\widetilde{Y}_i = Y_i - \bar{Y}_{g(i)}$. Note
that the sample covariance of a cluster-level variable (e.g $D_{g(i)}$) and a
within-cluster residual $\widetilde{Y}_i$ is zero.

\begin{equation}\label{eq:ovbderivunit}
\begin{aligned}
\hat{\tau}_{\text{res}}^{\text{U}}
&= \frac{\Cov\left(D_{g(i)}^{\perp \bu{X}}, Y_i^{\perp \bu{X}}\right)}
{\Var\left(D_{g(i)}^{\perp \bu{X}}\right)} \\
&= \frac{\Cov\left(D_{g(i)}^{\perp \bu{X}},
\widetilde{Y}_i^{\perp \bu{X}} + \bar{Y}_{g(i)}^{\perp \bu{X}}\right)}
{\Var\left(D_{g(i)}^{\perp \bu{X}}\right)} \\
&= \frac{\Cov\left(D_{g(i)}^{\perp \bu{X}},
\widetilde{Y}_i^{\perp \bu{X}} + \hat{\tau} D_{g(i)}^{\perp \bu{X}}
+ \hat{\gamma} Z_{g(i)}^{\perp \bu{X}}\right)}
{\Var\left(D_{g(i)}^{\perp \bu{X}}\right)} \\
&= \frac{\Cov\left(D_{g(i)}^{\perp \bu{X}},
\widetilde{Y}_i^{\perp \bu{X}}\right)}{\Var\left(D_{g(i)}^{\perp \bu{X}}\right)}
+ \hat{\tau}\frac{\Cov\left(D_{g(i)}^{\perp \bu{X}},
D_{g(i)}^{\perp \bu{X}}\right)}{\Var\left(D_{g(i)}^{\perp \bu{X}}\right)}
+ \hat{\gamma}\frac{\Cov\left(D_{g(i)}^{\perp \bu{X}},
Z_{g(i)}^{\perp \bu{X}}\right)}{\Var\left(D_{g(i)}^{\perp \bu{X}}\right)} \\
&= \hat{\tau} + \hat{\gamma}\hat{\theta}
\end{aligned}
\end{equation}

The last step follows from the equivalence of cluster-aggregated and individual
regressions under a balanced cluster design \citep{kloek1981ols, moulton1986random,
moulton1990illustration, wooldridge2003cluster}. Therefore
$\Cov\left(D_g^{\perp \bu{X}}, Z_g^{\perp \bu{X}}\right)
= \Cov\left(D_{g(i)}^{\perp \bu{X}}, Z_{g(i)}^{\perp \bu{X}}\right)$.\footnote{With
imbalanced clusters, the individual-level covariance equals to the $N_g$ weighted
group-level covariance, as in the standard group-/unit-level regression
equivalence.} The estimated treatment effects and the magnitude of omitted
variable bias are the same for the individual-level and the cluster-level
regression and $\hat{\tau}_{\text{res}}^{\text{U}} = \hat{\tau}_{\text{res}}$.

But directly applying the \citet{cinelli2020making} re-parameterization to the
\emph{unit-level} regression yields an expression in terms of \emph{unit-level}
partial $R^2$ values. The outcome-confounder sensitivity parameter is not
equivalent to the corresponding \emph{group-level} partial-$R^2$ as it
incorporates otherwise irrelevant variation that cannot be a source of
confounding: variation within-cluster.

\begin{equation}\label{eq:biasunit}
\left|\bias\right| = \sqrt{ R^2_{Y_i \sim Z_{g(i)} \mid D, \bu{X}}
\times \frac{R^2_{D_{g(i)} \sim Z_{g(i)} \mid \bu{X}}}
{1 - R^2_{D_{g(i)} \sim Z_{g(i)} \mid \bu{X}}} }
\times \frac{\Sd\left(Y_i^{\perp D, \bu{X}}\right)}
{\Sd\left(D_{g(i)}^{\perp \bu{X}}\right)}
\end{equation}

\begin{equation}\label{eq:relativebiasunit}
\text{relative bias} = \frac{\left| R_{Y_i \sim Z_{g(i)} \mid D, \bu{X}}
\times f_{D_{g(i)} \sim Z_{g(i)} \mid \bu{X}} \right|}
{\left| f_{Y_i \sim D_{g(i)} \mid \bu{X}} \right|}
\end{equation}

As before, it is possible to substitute
$R^2_{D_{g(i)} \sim Z_{g(i)} \mid \bu{X}} = R^2_{D_g \sim Z_g \mid \bu{X}}$ and
$\Sd\left(D_{g(i)}^{\perp \bu{X}}\right) = \Sd\left(D_g^{\perp \bu{X}}\right)$.
However, the partial-$R^2$ involving the unit-level outcome $Y_i$ includes both
the within-cluster and between-cluster variation. Only the latter can be
explained by a cluster-level confounder $Z$. As a result, the two $R^2$
parameters sit on different scales. While treatment-confounder partial $R^2$ can
plausibly range from $0$ to $1$, the outcome-confounder partial $R^2$ is
constrained to a quantity that can be written as the square of a Pearson
``correlation ratio'' $\eta^2$ \citep{pearson1905general, pearson1915partial,
kennedy1970eta, cohen1973eta}.

Let $\bu{G}$ denote the set of cluster indicators. Define
$\eta^2_{Y \mid D, \bu{X}} \equiv 1 -
\frac{\Var\left(Y_i^{\perp \bu{G}, D, \bu{X}}\right)}
{\Var\left(Y_i^{\perp D, \bu{X}}\right)} = R^2_{Y_i \sim \bu{G} \mid D, \bu{X}}$
as the share of the variation in the outcome that is explained by conditioning on
the cluster indicators beyond conditioning on the cluster-level treatment and
observed covariates. This is the ``non-parametric'' $R^2$ of
\citet{doksum1995nonparametric} and in the case where all covariates are measured
at the cluster level is equal to
$\frac{\Var\left(\bar{Y}_g^{\perp D, \bu{X}}\right)}
{\Var\left(\bar{Y}_g^{\perp D, \bu{X}}\right)
+ \Var\left(\widetilde{Y}_i^{\perp D, \bu{X}}\right)}$: the ratio of the
between-cluster residual variance to the total (between-cluster plus
within-cluster) residual variance. However, introducing individual-level
covariates complicates this interpretation slightly as shown later in
Section~\ref{sec:individualcovariates}. This $\eta^2$ is straightforward to
compute from an auxiliary regression of the residuals of
Equation~\ref{eq:unitrestricted} on the cluster dummies.

The unit-level outcome partial-$R^2$ is equal to the cluster-average partial
$R^2$ re-scaled by the squared partial-$\eta$. Using the definition of the
partial $R^2$ in terms of variance explained

\begin{equation}\label{eq:unittoclusterR2}
\begin{aligned}
R^2_{Y_i \sim Z_{g(i)} \mid D, \bu{X}}
&= 1 - \frac{\Var\left(Y_i^{\perp Z, D, \bu{X}}\right)}
{\Var\left(Y_i^{\perp D, \bu{X}}\right)} \\
&= 1 - \frac{\Var\left(\bar{Y}_g^{\perp Z, D, \bu{X}}\right)
+ \Var\left(\widetilde{Y}_i^{\perp Z, D, \bu{X}}\right)}
{\Var\left(\bar{Y}_g^{\perp D, \bu{X}}\right)
+ \Var\left(\widetilde{Y}_i^{\perp D, \bu{X}}\right)} \\
&= 1 - \frac{\Var\left(\bar{Y}_g^{\perp Z, D, \bu{X}}\right)
+ \Var\left(\widetilde{Y}_i\right)}
{\Var\left(\bar{Y}_g^{\perp D, \bu{X}}\right)
+ \Var\left(\widetilde{Y}_i\right)} \\
&= \frac{\Var\left(\bar{Y}_g^{\perp D, \bu{X}}\right)
- \Var\left(\bar{Y}_g^{\perp Z, D, \bu{X}}\right)}
{\Var\left(\bar{Y}_g^{\perp D, \bu{X}}\right)
+ \Var\left(\widetilde{Y}_i\right)} \\
&= \frac{\Var\left(\bar{Y}_g^{\perp D, \bu{X}}\right)
- \Var\left(\bar{Y}_g^{\perp Z, D, \bu{X}}\right)}
{\Var\left(\bar{Y}_g^{\perp D, \bu{X}}\right)
+ \Var\left(\widetilde{Y}_i^{\perp D, \bu{X}}\right)} \\
&= \frac{\Var\left(\bar{Y}_g^{\perp D, \bu{X}}\right)}
{\Var\left(\bar{Y}_g^{\perp D, \bu{X}}\right)
+ \Var\left(\widetilde{Y}_i^{\perp D, \bu{X}}\right)}
\times \left(1 - \frac{\Var\left(\bar{Y}_g^{\perp Z, D, \bu{X}}\right)}
{\Var\left(\bar{Y}_g^{\perp D, \bu{X}}\right)}\right) \\
&= \eta^2_{Y \mid D, \bu{X}} \times R^2_{\bar{Y}_g \sim Z_g \mid D, \bu{X}}
\end{aligned}
\end{equation}

Substituting back into the bias expression yields

\begin{equation}\label{eq:biasadjusted}
\begin{aligned}
\left|\bias\right|
&= \sqrt{ \eta^2_{Y \mid D, \bu{X}}
\times R^2_{\bar{Y}_g \sim Z_g \mid D, \bu{X}}
\times \frac{R^2_{D_g \sim Z_g \mid \bu{X}}}{1 - R^2_{D_g \sim Z_g \mid \bu{X}}} }
\times \frac{\Sd\left(Y_i^{\perp D, \bu{X}}\right)}
{\Sd\left(D_g^{\perp \bu{X}}\right)} \\
&= \sqrt{ \eta^2_{Y \mid D, \bu{X}}
\times R^2_{\bar{Y}_g \sim Z_g \mid D, \bu{X}}
\times \frac{R^2_{D_g \sim Z_g \mid \bu{X}}}{1 - R^2_{D_g \sim Z_g \mid \bu{X}}} }
\times \frac{\Sd\left(\bar{Y}_g^{\perp D, \bu{X}}\right)}
{\Sd\left(D_g^{\perp \bu{X}}\right)} \times \frac{1}{\eta_{Y \mid D, \bu{X}}}
\end{aligned}
\end{equation}

Therefore, rescaling from the bias factor returned by the unit-level regression
$\text{BF}^{\text{U}} = R_{Y_i \sim Z_{g(i)} \mid D_{g(i)}, \bu{X}}
\times f_{D_g \sim Z_g \mid \bu{X}}$ to the cluster-level bias factor requires
simply dividing by the partial correlation ratio of the outcome given $\bu{X}$
and $D$.

\begin{equation}\label{eq:unitclusterBF}
\begin{aligned}
\text{BF} &= \frac{1}{\eta_{Y \mid D, \bu{X}}} \times \text{BF}^{\text{U}}
\end{aligned}
\end{equation}

While it is still possible to conduct a sensitivity analysis using the unit-level
rather than cluster-level partial-$R^2$ values to generate plots of the bias
contours, interpreting the parameters substantively is more difficult. Moreover,
both of the proposed ``minimal reporting'' quantities in
\citet{cinelli2020making} are potentially misleading in this setting. The raw
robustness value computed from this regression $\text{RV}^{\text{U}}$ posits a
confounder that is ``equally strong'' on the treatment and outcome sides by
setting the unit level $R^2_{Y_i \sim Z_{g(i)} \mid D_{g(i)}, \bu{X}}$ equal to
the cluster-level $R^2_{D_{g(i)} \sim Z_{g(i)} \mid \bu{X}}$. However, as
Equation~\ref{eq:unittoclusterR2} shows, this is not an ``equal strength''
scenario.

Setting the unit-level outcome-confounder partial-$R^2$ equal to the
cluster-level treatment-confounder partial $R^2$ actually implies a confounder
that is $1/\eta^2_{Y \mid D, \bu{X}}$ times stronger with respect to the outcome
compared to treatment. Because the unit-level $R^2$ is contaminated by irrelevant
within-cluster variance, the analysis ascribes \emph{more} influence to the
confounder than it can actually have.
$R^2_{Y_i \sim Z_{g(i)} \mid D_{g(i)}, \bu{X}}$ and
$R^2_{D_{g(i)} \sim Z_{g(i)} \mid \bu{X}}$ are not on comparable scales.
Furthermore, the extreme scenario analysis considers a scenario that is so extreme
as to be \emph{impossible}: a cluster-level confounder that explains all of the
between and within-cluster variation. The correct upper bound on the unit-level
outcome-confounder sensitivity parameter is $\eta^2_{Y \mid D, \bu{X}}$, not $1$.

Luckily it is very simple to recover both of these summary statistics from the
cluster-level analysis using the reported \emph{unit-level} regression without
needing to actually fit a regression at the cluster-aggregated level. The
correction factor is identical across both minimal reporting quantities and
depends entirely on $\eta^2_{Y \mid D, \bu{X}}$. For the robustness value, it
suffices to simply correct the unit-level partial-$f$
$f_{Y_i \sim D_{g(i)} \mid \bu{X}}$ for the outcome on treatment conditional on
observed covariates such that it equals $f_{\bar{Y}_g \sim D_g \mid \bu{X}}$.

\begin{equation}\label{eq:unittoclusterf}
\begin{aligned}
f^2_{Y_i \sim D_{g(i)} \mid \bu{X}}
&= \frac{R^2_{Y_i \sim D_{g(i)} \mid \bu{X}}}
{1 - R^2_{Y_i \sim D_{g(i)} \mid \bu{X}}} \\
&= \frac{\Var\left(Y_i^{\perp \bu{X}}\right)
- \Var\left(Y_i^{\perp D, \bu{X}}\right)}
{\Var\left(Y_i^{\perp D, \bu{X}}\right)} \\
&= \frac{\Var\left(\bar{Y}_g^{\perp \bu{X}}\right)
+ \Var\left(\widetilde{Y}_i^{\perp \bu{X}}\right)
- \Var\left(\bar{Y}_g^{\perp D, \bu{X}}\right)
- \Var\left(\widetilde{Y}_i^{\perp D, \bu{X}}\right)}
{\Var\left(\bar{Y}_g^{\perp D, \bu{X}}\right)
+ \Var\left(\widetilde{Y}_i^{\perp D, \bu{X}}\right)} \\
&= \frac{\Var\left(\bar{Y}_g^{\perp \bu{X}}\right)
+ \Var\left(\widetilde{Y}_i\right)
- \Var\left(\bar{Y}_g^{\perp D, \bu{X}}\right)
- \Var\left(\widetilde{Y}_i^{\perp D, \bu{X}}\right)}
{\Var\left(\bar{Y}_g^{\perp D, \bu{X}}\right)
+ \Var\left(\widetilde{Y}_i\right)} \\
&= \frac{\Var\left(\bar{Y}_g^{\perp \bu{X}}\right)
- \Var\left(\bar{Y}_g^{\perp D, \bu{X}}\right)}
{\Var\left(\bar{Y}_g^{\perp D, \bu{X}}\right)
+ \Var\left(\widetilde{Y}_i\right)} \\
&= \frac{\Var\left(\bar{Y}_g^{\perp D, \bu{X}}\right)}
{\Var\left(\bar{Y}_g^{\perp D, \bu{X}}\right)
+ \Var\left(\widetilde{Y}_i^{\perp D, \bu{X}}\right)}
\times \frac{\Var\left(\bar{Y}_g^{\perp \bu{X}}\right)
- \Var\left(\bar{Y}_g^{\perp D, \bu{X}}\right)}
{\Var\left(\bar{Y}_g^{\perp D, \bu{X}}\right)} \\
&= \eta^2_{Y \mid D, \bu{X}} \times f^2_{\bar{Y}_g \sim D_g \mid \bu{X}}
\end{aligned}
\end{equation}

The adjusted robustness value can therefore be obtained by computing
$f_q = q\left| 1/\eta_{Y \mid D, \bu{X}} \times f_{Y_i \sim D_{g(i)} \mid \bu{X}}
\right|$ and substituting into Equation~\ref{eq:rvcluster}. Since
$\eta_{Y \mid D, \bu{X}} \leq 1$, this adjustment will always have the effect of
\emph{inflating} the robustness value (except in the edge case where there is no
within-cluster variation in outcome)

\begin{equation}\label{eq:rvexact}
\frac{\text{RV}_q^{\text{U}}}{\sqrt{1 - \text{RV}_q^{\text{U}}}}
= \eta_{Y \mid D, \bu{X}} \times \frac{\text{RV}_q}{\sqrt{1 - \text{RV}_q}}
\end{equation}

An identical correction can be applied to the extreme scenario analysis and
follows from the relationship between the cluster-level and unit-level bias
factors (Equation~\ref{eq:unitclusterBF}). Setting
$R_{Y_i \sim Z_{g(i)} \mid D, \bu{X}}$ in Equation~\ref{eq:relativebiasunit} to
its largest possible value, $\eta_{Y \mid D, \bu{X}}$ and solving for the value of
$f_{D_{g(i)} \sim Z_{g(i)} \mid \bu{X}}$ needed to set the relative bias to $1$
yields $1/\eta_{Y \mid D, \bu{X}} \times f_{Y_i \sim D_{g(i)} \mid \bu{X}}$. As
with Equation~\ref{eq:rvexact}, this re-scales the extreme robustness value as:

\begin{equation}\label{eq:xrvexact}
\frac{\text{XRV}_q^{\text{U}}}{1 - \text{XRV}_q^{\text{U}}}
= \eta^2_{Y \mid D, \bu{X}} \times \frac{\text{XRV}_q}{1 - \text{XRV}_q}
\end{equation}

\section{Benchmarking with clustered treatment assignment and the problems of
individual-level covariates}
\label{sec:individualcovariates}

While the reparameterization of the omitted variables bias into scale-invariant
$R^2$ terms provides some improvement in interpretability, researchers still
often have difficulty in reasoning about these quantities directly. In some
settings, theory might suggest many hypothetical confounders that could reach the
necessary treatment and outcome $R^2$ thresholds to overturn the findings. In
others, researchers might believe that the number of potential threats to
identification is limited. Researchers typically have a sense of which confounders
are most relevant and have already attempted to control for them in $\bu{X}$.
Therefore, a common sensitivity analysis question is to ask whether a hypothetical
confounder \emph{as strong as} an observed confounder already accounted for would
be enough to overturn the results.

\citet{cinelli2020making} introduces a bounding approach to compare the
partial-$R^2$ of the unobserved confounder to the partial-$R^2$ of some set of
observed confounders, conditional on treatment and the remaining included
covariates. Let $\bu{X}_j$ denote the set of observed confounders against which a
researcher wishes to benchmark and $\bu{X}_{-j}$ the remaining covariates.
\citet{cinelli2020making} defines the relative strength of the omitted confounder
relative to the observed confounders with respect to the treatment $k_D$ and
outcome $k_Y$. For the cluster-level regression, write these as:

\begin{equation}\label{eq:kdky}
\begin{aligned}
k_D &\equiv \frac{R^2_{D_g \sim Z_g \mid \bu{X}_{-j}}}
{R^2_{D_g \sim \bu{X}_j \mid \bu{X}_{-j}}} \\
k_Y &\equiv \frac{R^2_{\bar{Y}_g \sim Z_g \mid D, \bu{X}_{-j}}}
{R^2_{\bar{Y}_g \sim \bu{X}_j \mid D, \bu{X}_{-j}}}
\end{aligned}
\end{equation}

Given researcher choices of $k_D$ and $k_Y$, which capture the notion of a
confounder $k_D$ times as strong as $\bu{X}_j$ with respect to treatment and $k_Y$
times as strong with respect to outcome, \citet{cinelli2020making} shows that the
omitted confounder strength can be bounded by

\begin{equation}\label{eq:benchmarkcluster}
\begin{aligned}
R^2_{D_g \sim Z_g \mid \bu{X}}
&= k_D \times f^2_{D_g \sim \bu{X}_j \mid \bu{X}_{-j}} \\
R^2_{\bar{Y}_g \sim Z_g \mid D, \bu{X}}
&\leq \lambda^2 \times f^2_{\bar{Y}_g \sim \bu{X}_j \mid D, \bu{X}_{-j}}
\end{aligned}
\end{equation}

where $\lambda \equiv \frac{\sqrt{k_Y}
+ \left| f_{k_D} \times f_{D_g \sim \bu{X}_j \mid \bu{X}_{-j}} \right|}
{\sqrt{1 - f^2_{k_D} \times f^2_{D_g \sim \bu{X}_j \mid \bu{X}_{-j}}}}$ and
$f^2_{k_D} \equiv \frac{k_D \times R^2_{D_g \sim \bu{X}_j \mid \bu{X}_{-j}}}
{1 - k_D \times R^2_{D_g \sim \bu{X}_j \mid \bu{X}_{-j}}}$.

The bounds therefore only require
$f^2_{\bar{Y}_g \sim \bu{X}_j \mid D, \bu{X}_{-j}}$ and
$R^2_{D_g \sim \bu{X}_j \mid \bu{X}_{-j}}$. The former can be read directly from
the outcome regression while the latter requires an auxiliary regression of the
treatment indicator on the benchmark covariates after partialling out the
remaining observed confounders.

What happens when one instead fits the unit-level regression rather than the
cluster-level regression? Since the treatment-side partial-$R^2$ are equivalent
with a cluster-level confounder,
$R^2_{D_{g(i)} \sim Z_g \mid \bu{X}} = R^2_{D_g \sim Z_g \mid \bu{X}}$. Likewise,
$\lambda$ remains unchanged as it is composed of researcher-specified constants
and regressions involving only cluster-level variables. However, the bound on the
unit-level outcome-confounder partial-$R^2$ is

\begin{equation}\label{eq:benchmarkunit}
R^2_{Y_i \sim Z_{g(i)} \mid D, \bu{X}}
\leq \lambda^2 \times f^2_{Y_i \sim \bu{X}_j \mid D, \bu{X}_{-j}}
\end{equation}

From Equation~\ref{eq:unittoclusterf},
$f^2_{Y_i \sim \bu{X}_j \mid D, \bu{X}_{-j}}
= \eta^2_{Y \mid D, \bu{X}} \times f^2_{\bar{Y}_g \sim \bu{X}_j \mid D, \bu{X}_{-j}}$
and from Equation~\ref{eq:unittoclusterR2},
$R^2_{Y_i \sim Z_{g(i)} \mid D, \bu{X}}
= \eta^2_{Y \mid D, \bu{X}} \times R^2_{\bar{Y}_g \sim Z_g \mid D, \bu{X}}$. The
bound is deflated by exactly the same factor that deflates the partial-$R^2$ from
the group-level regression. Therefore, for a cluster-level covariate, applying the
bounding exercise yields the same adjusted effect estimate for a given $k_D$ and
$k_Y$ under either the unit-level or the cluster-level regression.

But one of the primary reasons researchers fit the unit-level regression is that
it permits easy inclusion of covariates measured for each unit $i$. However, the
purpose of this is often unclear. Individual-level factors under the clustered
observational design can act as confounders only through their cluster-level
aggregates. The covariance of the within-cluster residual with any cluster-level
confounder is $0$. Moreover, the within-cluster variation in these covariates net
of their cluster-level averages contributes nothing to the estimation of the
treatment effect $\hat{\tau}$ and when clustering standard errors, nothing to
variance estimation \citep{liang1986longitudinal}. Despite this, it is common to
see regressions that include a mix of covariates measured at both the cluster and
the unit level.

I consider two ways of adding an individual-level covariate denoted $W_i$ to the
``restricted'' regression specification in Equation~\ref{eq:unitrestricted}. The
first, following \citet{mundlak1978pooling}, involves including the cluster-level
mean $\bar{W}_g$ and the within-cluster deviation
$\widetilde{W}_i = W_i - \bar{W}_{g(i)}$ as separate regressors.\footnote{By
Frisch-Waugh-Lovell, it is equivalent to just include $\bar{W}_g$ alongside
$W_i$}

\begin{equation}\label{eq:mundlakspec}
Y_i = \hat{\tau}_{\text{res}}^{\text{M}} D_{g(i)}
+ \hat{\omega}^{\text{M}}_{\text{res}} \widetilde{W}_i
+ \hat{\alpha}^{\text{M}}_{\text{res}} \bar{W}_{g(i)}
+ X_{g(i)}^T \hat{\beta}^{\text{M}}_{\text{res}}
+ \hat{\epsilon}^{\text{res},\text{M}}
\end{equation}

$\hat{\tau}_{\text{res}}^{\text{M}}$ is unchanged if one omits $\widetilde{W}_i$
from the regression as it is orthogonal to all other cluster-level regressors by
construction. By Frisch-Waugh-Lovell

\begin{equation}
\begin{aligned}
\hat{\tau}_{\text{res}}^{\text{M}}
&= \frac{\Cov\left(D_{g(i)}^{\perp \widetilde{W}, \bar{W}, \bu{X}},
Y_i^{\perp \widetilde{W}, \bar{W}, \bu{X}}\right)}
{\Var\left(D_{g(i)}^{\perp \widetilde{W}, \bar{W}, \bu{X}}\right)} \\
&= \frac{\Cov\left(D_{g(i)}^{\perp \widetilde{W}, \bar{W}, \bu{X}},
\widetilde{Y}_i^{\perp \widetilde{W}, \bar{W}, \bu{X}}
+ \bar{Y}_{g(i)}^{\perp \widetilde{W}, \bar{W}, \bu{X}}\right)}
{\Var\left(D_{g(i)}^{\perp \widetilde{W}, \bar{W}, \bu{X}}\right)} \\
&= \frac{\Cov\left(D_{g(i)}^{\perp \bar{W}, \bu{X}},
\widetilde{Y}_i^{\perp \widetilde{W}}
+ \bar{Y}_{g(i)}^{\perp \bar{W}, \bu{X}}\right)}
{\Var\left(D_{g(i)}^{\perp \bar{W}, \bu{X}}\right)} \\
&= \frac{\Cov\left(D_{g(i)}^{\perp \bar{W}, \bu{X}},
\widetilde{Y}_i^{\perp \widetilde{W}}\right)}
{\Var\left(D_{g(i)}^{\perp \bar{W}, \bu{X}}\right)}
+ \frac{\Cov\left(D_{g(i)}^{\perp \bar{W}, \bu{X}},
\bar{Y}_{g(i)}^{\perp \bar{W}, \bu{X}}\right)}
{\Var\left(D_{g(i)}^{\perp \bar{W}, \bu{X}}\right)} \\
&= \frac{\Cov\left(D_{g(i)}^{\perp \bar{W}, \bu{X}},
\bar{Y}_{g(i)}^{\perp \bar{W}, \bu{X}}\right)}
{\Var\left(D_{g(i)}^{\perp \bar{W}, \bu{X}}\right)}
\end{aligned}
\end{equation}

Omitting the within-unit variation $\widetilde{W}_i$ also yields a specification
equivalent to Equation~\ref{eq:unitrestricted} in which all covariates, including
$\bar{W}_g$, vary only between clusters. Therefore, all of the prior results apply
cleanly to including cluster-level averages among the set of unit-level covariates
irrespective of whether the regression does or does not also include
$\widetilde{W}_i$. In fact, the adjusted robustness values and extreme scenario
analysis remain the same even though
$\eta^2_{Y \mid D, \widetilde{W}, \bar{W}, \bu{X}}
\neq \eta^2_{Y \mid D, \bar{W}, \bu{X}}$. Adding $\widetilde{W}_i$ scales $\eta^2$
and the unit-level partial-$f^2$ by a common factor, leaving the ratio that enters
into $\text{RV}_q$ and the extreme scenario analysis unchanged.

\begin{equation}\label{eq:mundlakinvariance}
\begin{aligned}
\eta^2_{Y \mid D, \widetilde{W}, \bar{W}, \bu{X}}
&= \frac{1}{1 - R^2_{Y_i \sim \widetilde{W}_i \mid D, \bar{W}, \bu{X}}}
\times \eta^2_{Y \mid D, \bar{W}, \bu{X}} \\
f^2_{Y_i \sim D_{g(i)} \mid \widetilde{W}, \bar{W}, \bu{X}}
&= \frac{1}{1 - R^2_{Y_i \sim \widetilde{W}_i \mid D, \bar{W}, \bu{X}}}
\times f^2_{Y_i \sim D_{g(i)} \mid \bar{W}, \bu{X}}
\end{aligned}
\end{equation}

In applied practice, the \citet{mundlak1978pooling} within-between specification
is unfortunately rare. Instead, researchers will typically fit a regression
including only $W_i$ without the group-level means.

\begin{equation}\label{eq:pooledspec}
Y_i = \hat{\tau}_{\text{res}}^{\text{pool}} D_{g(i)}
+ \hat{\alpha}_{\text{res}}^{\text{pool}} W_i
+ X_{g(i)}^T \hat{\beta}^{\text{pool}}_{\text{res}}
+ \hat{\epsilon}_i^{\text{res},\text{pool}}
\end{equation}

This is equivalent to imposing an additional constraint on
Equation~\ref{eq:mundlakspec} that the coefficients on the between variation and
the within variation are equivalent
$\hat{\omega}_{\text{res}}^{\text{M}} = \hat{\alpha}_{\text{res}}^{\text{M}}$.
\citet{mundlak1978pooling} shows that the pooled coefficient
$\hat{\alpha}_{\text{res}}^{\text{pool}}$ is the variance-weighted average of the
coefficients from the unconstrained regression

\begin{equation}
\begin{aligned}
\hat{\alpha}_{\text{res}}^{\text{pool}}
&= \frac{\Var\left(\bar{W}_g^{\perp D, \bu{X}}\right)
\hat{\alpha}_{\text{res}}^{\text{M}}
+ \Var\left(\widetilde{W}_i\right) \hat{\omega}_{\text{res}}^{\text{M}}}
{\Var\left(W_i^{\perp D, \bu{X}}\right)} \\
&= \eta^2_{W \mid D, \bu{X}} \times \hat{\alpha}_{\text{res}}^{\text{M}}
+ (1 - \eta^2_{W \mid D, \bu{X}}) \times \hat{\omega}_{\text{res}}^{\text{M}}
\end{aligned}
\end{equation}

where $\eta^2_{W \mid D, \bu{X}}$ is now the within-between partial correlation
ratio for covariate $W$. This affects the sensitivity analysis in two ways. First,
it leaves an additional source of variation for an unobserved confounder $Z_g$ to
explain, the mis-specification resulting from adjusting for $W_i$ rather than
$\bar{W}_g$. Including the individual-level covariate into the adjustment set for
$\eta^2_{Y \mid D, W, \bu{X}}$ means that it now encompasses both the
between-within cluster variation in the outcome as well as a term reflecting the
discrepancy between the pooled coefficient
$\hat{\alpha}_{\text{res}}^{\text{pool}}$ and the coefficient on the
within-variation $\hat{\omega}^{\text{M}}_{\text{res}}$ of $W$ from the Mundlak
regression.

\begin{equation}\label{eq:ceilingpoolvar}
\begin{aligned}
\eta^2_{Y \mid D, W, \bu{X}}
&= 1 - \frac{\Var\left(Y_i^{\perp \bu{G}, D, W, \bu{X}}\right)}
{\Var\left(Y_i^{\perp D, W, \bu{X}}\right)} \\
&= 1 - \frac{\Var\left(\widetilde{Y}_i^{\perp W}\right)}
{\Var\left(Y_i^{\perp D, W, \bu{X}}\right)} \\
&= \frac{\Var\left(\bar{Y}_g^{\perp D, W, \bu{X}}\right)
+ \Var\left(\widetilde{Y}_i^{\perp D, W, \bu{X}}\right)}
{\Var\left(Y_i^{\perp D, W, \bu{X}}\right)}
- \frac{\Var\left(\widetilde{Y}_i^{\perp W}\right)}
{\Var\left(Y_i^{\perp D, W, \bu{X}}\right)} \\
&= \frac{\Var\left(\bar{Y}_g^{\perp D, W, \bu{X}}\right)}
{\Var\left(Y_i^{\perp D, W, \bu{X}}\right)}
+ \frac{\Var\left(\widetilde{Y}_i^{\perp D, W, \bu{X}}\right)
- \Var\left(\widetilde{Y}_i^{\perp W}\right)}
{\Var\left(Y_i^{\perp D, W, \bu{X}}\right)}
\end{aligned}
\end{equation}

Because $\hat{\omega}^{\text{M}}_{\text{res}}$ is the least-squares slope of
$\widetilde{Y}_i$ on $\widetilde{W}_i$ whereas Equation~\ref{eq:pooledspec} fixes
the within-cluster variation to $\hat{\alpha}^{\text{pool}}_{\text{res}}$, the
within-cluster residual variances differ by
$\Var\left(\widetilde{Y}_i^{\perp D, W, \bu{X}}\right)
- \Var\left(\widetilde{Y}_i^{\perp W}\right)
= (\hat{\alpha}^{\text{pool}}_{\text{res}}
- \hat{\omega}^{\text{M}}_{\text{res}})^2 \Var\left(\widetilde{W}_i\right)$.
Substituting into Equation~\ref{eq:ceilingpoolvar},

\begin{equation}\label{eq:etagap}
\eta^2_{Y \mid D, W, \bu{X}}
= \underbrace{\frac{\Var\left(\bar{Y}_g^{\perp D, W, \bu{X}}\right)}
{\Var\left(Y_i^{\perp D, W, \bu{X}}\right)}}_{\text{between-cluster share}}
+ \underbrace{\frac{(\hat{\alpha}^{\text{pool}}_{\text{res}}
- \hat{\omega}^{\text{M}}_{\text{res}})^2 \Var\left(\widetilde{W}_i\right)}
{\Var\left(Y_i^{\perp D, W, \bu{X}}\right)}}_{\text{Mundlak discrepancy}}
\end{equation}

Second, benchmarking against the raw unit-level confounder $W_i$ as opposed to the
cluster mean $\bar{W}_g$ can yield potentially misleading conclusions regarding
the strength of the confounder. Suppose a researcher fitting
Equation~\ref{eq:pooledspec} considers a confounder $k_D$ times as strong as $W_i$
in explaining treatment and $k_Y$ times as strong as $W_i$ in explaining the
outcome. How does this compare to the strength of confounding if instead the
researcher had benchmarked against the relevant confounder $\bar{W}_g$? Denote
these factors $k^*_D$ and $k^*_Y$ and write the ratios

\begin{equation}\label{eq:kratios}
\begin{aligned}
\frac{k^*_D}{k_D} &= \frac{R^2_{D_{g(i)} \sim W_i \mid \bu{X}}}
{R^2_{D_g \sim \bar{W}_g \mid \bu{X}}} \\
\frac{k^*_Y}{k_Y} &= \frac{R^2_{Y_i \sim W_i \mid D, \bu{X}}}
{R^2_{Y_i \sim \bar{W}_{g(i)} \mid D, \bu{X}}}
\end{aligned}
\end{equation}

The treatment-side partial $R^2$ follows from an argument similar to
Equation~\ref{eq:unittoclusterR2} as the only variance in $W$ that explains
treatment is the between-cluster variance

\begin{equation}\label{eq:treatmentbenchmarkscale}
R^2_{D_{g(i)} \sim W_i \mid \bu{X}}
= \eta^2_{W \mid \bu{X}} \times R^2_{D_g \sim \bar{W}_g \mid \bu{X}}
\end{equation}

The outcome-confounder relationship is a bit more complex as it involves
$\eta^2_{W \mid D, \bu{X}}$ and a covariance ratio term. Writing the partial-$R^2$
values as squared correlation coefficients:

\begin{equation}\label{eq:outcomebenchmarkscale}
\begin{aligned}
\frac{R^2_{Y_i \sim W_i \mid D, \bu{X}}}
{R^2_{Y_i \sim \bar{W}_{g(i)} \mid D, \bu{X}}}
&= \frac{\Cov\left(Y_i^{\perp D, \bu{X}}, W_i^{\perp D, \bu{X}}\right)^2}
{\Var\left(Y_i^{\perp D, \bu{X}}\right) \Var\left(W_i^{\perp D, \bu{X}}\right)}
\times \frac{\Var\left(Y_i^{\perp D, \bu{X}}\right)
\Var\left(\bar{W}_{g(i)}^{\perp D, \bu{X}}\right)}
{\Cov\left(Y_i^{\perp D, \bu{X}}, \bar{W}_{g(i)}^{\perp D, \bu{X}}\right)^2} \\
&= \left(\frac{\Cov\left(Y_i^{\perp D, \bu{X}}, W_i^{\perp D, \bu{X}}\right)}
{\Cov\left(Y_i^{\perp D, \bu{X}}, \bar{W}_{g(i)}^{\perp D, \bu{X}}\right)}\right)^2
\times \eta^2_{W \mid D, \bu{X}} \\
&= \left(\frac{\Cov\left(\bar{Y}_g^{\perp D, \bu{X}}, \bar{W}_g^{\perp D, \bu{X}}\right)
+ \Cov\left(\widetilde{Y}_i, \widetilde{W}_i\right)}
{\Cov\left(\bar{Y}_g^{\perp D, \bu{X}}, \bar{W}_g^{\perp D, \bu{X}}\right)}\right)^2
\times \eta^2_{W \mid D, \bu{X}} \\
&= \left(1 + \frac{\Cov\left(\widetilde{Y}_i, \widetilde{W}_i\right)}
{\Cov\left(\bar{Y}_g^{\perp D, \bu{X}}, \bar{W}_g^{\perp D, \bu{X}}\right)}\right)^2
\times \eta^2_{W \mid D, \bu{X}}
\end{aligned}
\end{equation}

Intuitively, the cluster means of the covariate can only covary with the cluster
means of the outcome, while the individual-level covariate covaries with both.
Substituting both into the expression for the ratios of $k$ yields

\begin{equation}\label{eq:kratios-factors}
\begin{aligned}
\frac{k_D^*}{k_D} &= \eta^2_{W \mid \bu{X}} \\
\frac{k_Y^*}{k_Y} &= \left(1 + \frac{\Cov\left(\widetilde{Y}_i, \widetilde{W}_i\right)}
{\Cov\left(\bar{Y}_g^{\perp D, \bu{X}}, \bar{W}_g^{\perp D, \bu{X}}\right)}\right)^2
\times \eta^2_{W \mid D, \bu{X}}
\end{aligned}
\end{equation}

On the treatment side, positing a hypothetical confounder ``as strong as'' $W_i$
is equivalent to assuming a confounder that is only $\eta^2_{W \mid \bu{X}}$ times
as strong as $\bar{W}_g$. Benchmarking against $W_i$ as opposed to the group-level
averages always yields a ``weaker'' test with respect to the treatment side
relationship. On the outcome side, benchmarking against $W_i$ could be
\emph{either} a ``stronger'' or ``weaker'' test, depending on within-cluster
covariance of the covariate and outcome. In the simplest case where the covariate
is irrelevant within cluster and only relevant between cluster, the benchmark is
weaker on both dimensions. Evaluating against a confounder as strong as $W_i$ is
anti-conservative compared to evaluating against $\bar{W}_g$. However, if $W_i$
explains a substantial amount of within-cluster variation in $Y_i$, treating this
as ``confounding strength'' on the outcome side may offset the treatment side
deflation and yield an overly severe test of robustness compared to just
benchmarking versus the group means.

\section{Replication: \texorpdfstring{\citet{reny2026rising}}{Reny et al. (2026)}
on climate vulnerability}
\label{sec:replication}

I illustrate the problems with the unit-level sensitivity analysis through a
re-analysis of \citet{reny2026rising}, which studies the effects of individuals'
vulnerability to risks from climate change on their attitudes towards climate
mitigation policies. The study focuses specifically on U.S. residents' exposure
to sea level rise, which they measure using a continuous risk score calculated at
the \emph{county level}. Data on climate attitudes was collected via a large-N
national online survey of adult U.S. residents who were geo-located to their
respective counties of residence. The primary analysis is an individual-level
regression of the outcome of interest, an eight-item scale of support for climate
mitigation policies, on the county-level risk score and observed covariates.
Although the survey contains $N = 2845$ respondents, they are nested within
$C = 932$ total counties with variation in the number of respondents per county.
Reported standard errors are clustered at the county level (CR2 correction).

The regression contains no cluster-level confounders and includes nine unit-level
covariates measured in the survey directly in the regression model (as in
Equation~\ref{eq:pooledspec}). These covariates are party identification,
conservative ideology, age, gender, college, race/ethnicity and income level (3
dummy indicators). The study estimates a positive and statistically significant
effect, finding that a one unit increase in exposure to sea level rise increases a
respondent's support for climate mitigation policies by $0.095$ points (SE =
$0.029$) on average. Appendix C4 of \citet{reny2026rising} presents the standard
reporting output from \texttt{sensemakr} applied to this regression. I reproduce
the relevant features of this table in Table~\ref{tab:c4}.\footnote{I omit the
standard errors/t-statistics and quantile robustness statistics that are default
in the \texttt{sensemakr} output and presented in Appendix C4 as these are
computed under the classical OLS assumptions rather than cluster-robust standard
errors presented in the main text.}

\begin{table}[H]
\centering
\begin{tabular}{lccc}
\toprule
\multicolumn{4}{c}{\emph{Outcome: policy scale}} \\
\midrule
\textbf{Treatment:} & \textbf{Est.}
& $\boldsymbol{R^2_{Y \sim D \mid \mathbf{X}}}$
& $\boldsymbol{\mathrm{RV}_{q=1}}$ \\
\midrule
sea level rise & 0.095 & 0.6\% & 7.4\% \\
\bottomrule
\multicolumn{4}{l}{\footnotesize Bound (1x pid7\_r):
$R^2_{Y \sim Z \mid D, \mathbf{X}} = 6.6\%$,
$R^2_{D \sim Z \mid \mathbf{X}} = 0.5\%$} \\
\end{tabular}
\caption{Replication of Appendix table C4 in \citet{reny2026rising}}
\label{tab:c4}
\end{table}

The original sensitivity analysis concludes that a confounder would have to
explain $7.4\%$ of the variation in both the outcome and the treatment in order to
drive the estimate to zero. Benchmarking against the included party ID scale
suggests that a confounder of comparable strength would be insufficient to break
the result. Notably, while party ID is fairly well correlated with the outcome,
its correlation with treatment is quite minimal. A comparable confounder on the
treatment side would need a significantly larger correlation with the outcome to
explain away the results. The extreme sensitivity analysis is a much harder test:
a confounder that explained $100\%$ of the outcome variance would need to only
explain .6\% of the variation in treatment to eliminate the result. This is very
close to the party ID benchmark.

But both of these statistics incorporate irrelevant within-cluster variation in
the confounders. While one might be concerned that coastal and inland counties
vary in the overall \emph{composition} of respondents (e.g. if coastal counties
are less Republican on average), the intra-cluster correlation between partisanship
and outcome poses no omitted variable bias threat. Therefore the reported
sensitivity statistics in Table~\ref{tab:c4} need to be adjusted by
$\eta^2_{Y \mid D, W}$. Calculating this quantity for the original regression
specification (with individual-level covariates) yields an adjustment factor of
$1/\eta_{Y \mid D, W} \approx 1.77$. Applying this to
Equation~\ref{eq:unittoclusterf} yields a cluster-adjusted robustness value of
$\text{RV}^{\text{C}}_1 = 12.7\%$. A county-level confounder must explain about
$12.7\%$ of the between-county variation in both exposure to sea level rise and
policy support to drive the estimate to zero. The results are nearly twice as
``robust'' as originally reported.

\begin{figure}[H]
\centering
\includegraphics[width=0.92\linewidth]{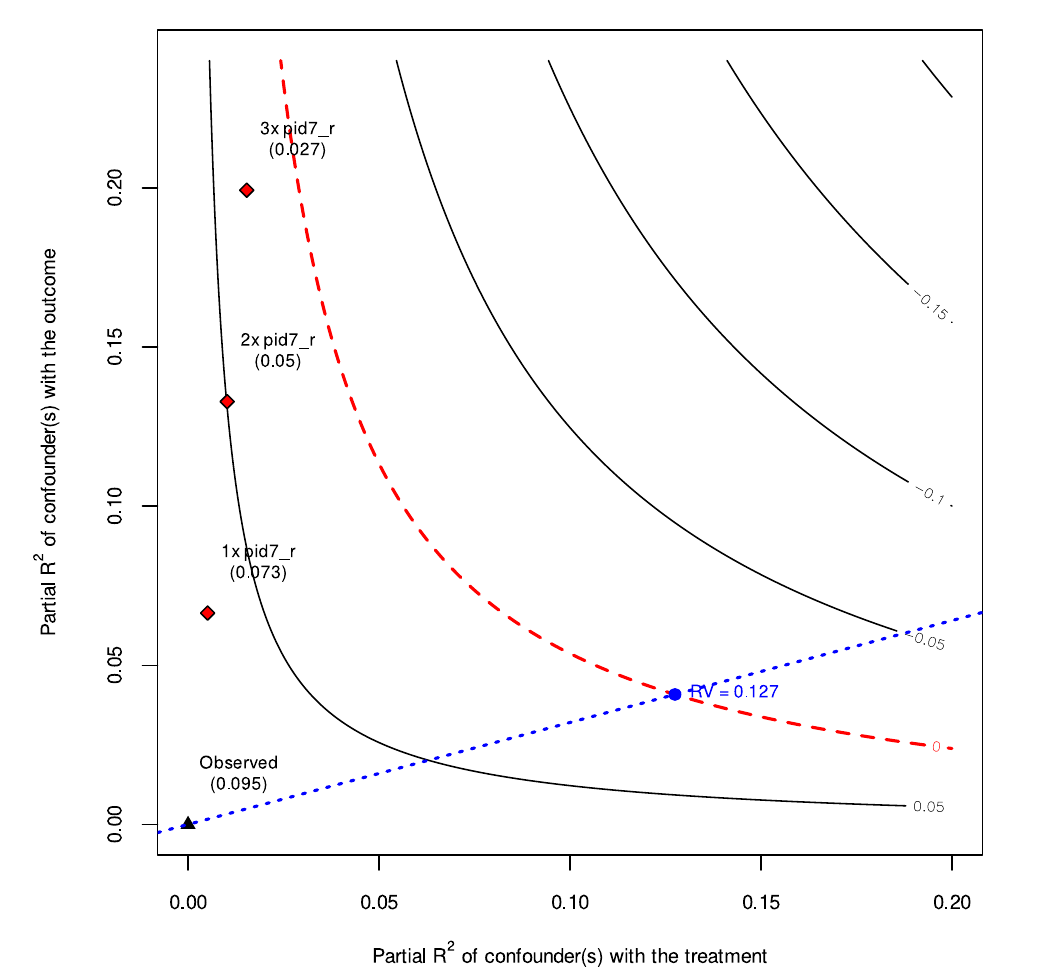}
\caption{Sensitivity contour plot for main results of \citet{reny2026rising}}
\label{fig:contour}
{\par\small\emph{Note:} Dashed blue line denotes the equiconfounding line for a
cluster-level confounder. It intersects the red zero-effect contour at the
adjusted robustness value for $q=1$.}
\end{figure}

Figure~\ref{fig:contour} displays the conventional contour plot of the original
sensitivity analysis along with the author's chosen covariate benchmark. I overlay
a dashed blue line to denote the true ``equiconfounding'' line implied by the
estimated adjustment for the between-cluster variation. The slope of this line is
equal to $\eta^2_{Y \mid D, W}$ and it intersects the red line denoting zero
effect at exactly $\text{RV}^{\text{C}}_1$.

\begin{figure}[H]
\centering
\includegraphics[width=0.92\linewidth]{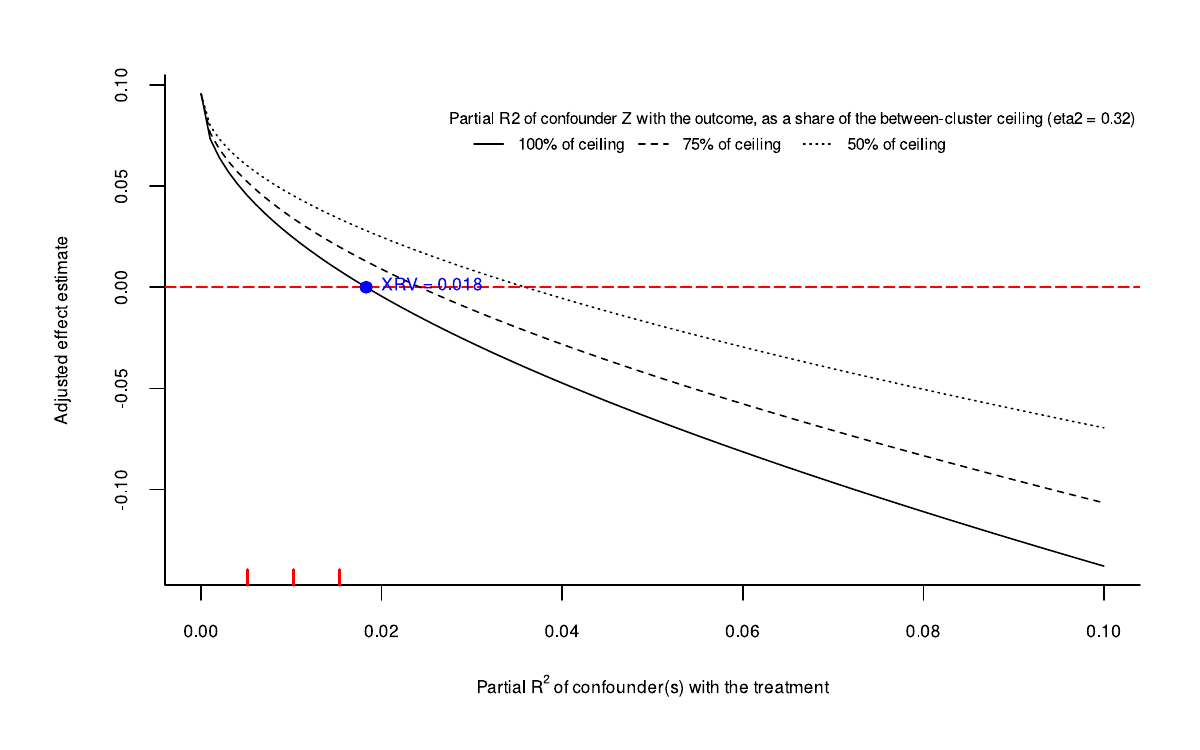}
\caption{Extreme scenario analysis for main results of \citet{reny2026rising}}
\label{fig:extreme}
\end{figure}

Figure~\ref{fig:extreme} applies the same correction to the extreme robustness
value to reflect the fact that a confounder cannot explain 100\% of the outcome
variance, it can explain at most $100\% \times \eta^2_{Y \mid D, W} = 32\%$ of it.
This increases the necessary partial-$R^2$ between treatment and outcome from
$0.59\%$ -- alarmingly close to the party ID benchmark $0.51\%$ -- to $1.83\%$.

But as Section~\ref{sec:individualcovariates} illustrates, benchmarking against
the raw individual covariates can be very misleading. I re-estimate the unit-level
regression from the original analysis including only the cluster-level averages of
each covariate. In addition, I estimate the equivalent cluster-level regression of
$\bar{Y}_g$ on the cluster-averaged covariates, weighted by cluster size to
illustrate the equivalence of these two approaches.

\begin{table}[H]
\centering
\begin{tabular}{lccccccc}
\toprule
\multicolumn{3}{c}{\emph{Outcome: policy scale}}
& \multicolumn{2}{c}{\textbf{Naive (unit level)}}
& \multicolumn{3}{c}{\textbf{Cluster-adjusted}} \\
\cmidrule(lr){1-3}\cmidrule(lr){4-5}\cmidrule(lr){6-8}
\textbf{Treatment:} & \textbf{Est.} & \textbf{S.E.}
& $\boldsymbol{R^2_{Y \sim D \mid \mathbf{X}}}$
& $\boldsymbol{\mathrm{RV}^{\mathrm{U}}_{q=1}}$
& $\boldsymbol{\eta^2_{Y \mid D, \mathbf{X}}}$
& $\boldsymbol{\mathrm{XRV}^{\mathrm{C}}_{q=1}}$
& $\boldsymbol{\mathrm{RV}^{\mathrm{C}}_{q=1}}$ \\
\midrule
sea level rise & 0.0837 & 0.0275 & 0.36\% & 5.8\% & 0.27 & 1.33\% & 10.9\% \\
\bottomrule
\multicolumn{8}{l}{\footnotesize $N = 2845$ respondents, $C = 932$ counties. CR2
standard error.} \\
\multicolumn{8}{l}{\footnotesize Bound (1x county-mean pid7\_r), cluster scaled:
$R^2_{Y \sim Z \mid \mathbf{X}, D} = 5.85\%$,
$R^2_{D \sim Z \mid \mathbf{X}} = 0.50\%$} \\
\end{tabular}
\caption{Unit-level regression of support for climate mitigation on county sea
level rise and covariate county means.}
\label{tab:aggunit}
\end{table}

Table~\ref{tab:aggunit} presents the estimated treatment effect and robustness
summaries from the unit-level regression. Adjusting for the county-level means
attenuates the point estimate slightly, suggesting that some of the omitted
variables bias was not being fully accounted for in the original analysis.
Nevertheless, the treatment effect remains positive ($0.0837$) and statistically
significant (robust SE $0.0275$).

A naive application of the robustness value to this regression would conclude that
a confounder that explains only $5.8\%$ of the variance in treatment and outcome
would break the result. However, only $27\%$ of the variation in the residual
outcome is between county. Most of what that hypothetical confounder would need to
explain is entirely irrelevant variation! Applying the correction
$1/\eta^2_{Y \mid D, \bu{X}}$ yields an adjusted robustness value of $10.9\%$ --
smaller than in the original regression but still sizeable. Moreover, the
benchmarking results remain essentially the same - a confounder that is as strong
as party ID would fail to eliminate the results. Party ID remains strongly
predictive of the outcome but surprisingly not well correlated with exposure to
sea level rise.

\begin{table}[H]
\centering
\begin{tabular}{lcccc}
\toprule
\multicolumn{5}{c}{\emph{Outcome: county-mean policy scale}} \\
\midrule
\textbf{Treatment:} & \textbf{Est.} & \textbf{S.E.}
& $\boldsymbol{R^2_{Y \sim D \mid \mathbf{X}}}$
& $\boldsymbol{\mathrm{RV}_{q=1}}$ \\
\midrule
sea level rise & 0.0837 & 0.0275 & 1.33\% & 10.9\% \\
\bottomrule
\multicolumn{5}{l}{\footnotesize $C = 932$ counties. HC2 standard error. Weighted
by number of units in county} \\
\multicolumn{5}{l}{\footnotesize Bound (1x county-mean pid7\_r):
$R^2_{Y \sim Z \mid D, \mathbf{X}} = 5.85\%$,
$R^2_{D \sim Z \mid \mathbf{X}} = 0.50\%$} \\
\end{tabular}
\caption{County-level regression of county-average support for climate mitigation
on county sea level rise and covariate county means.}
\label{tab:aggcounty}
\end{table}

\begin{figure}[H]
\centering
\includegraphics[width=0.49\linewidth]{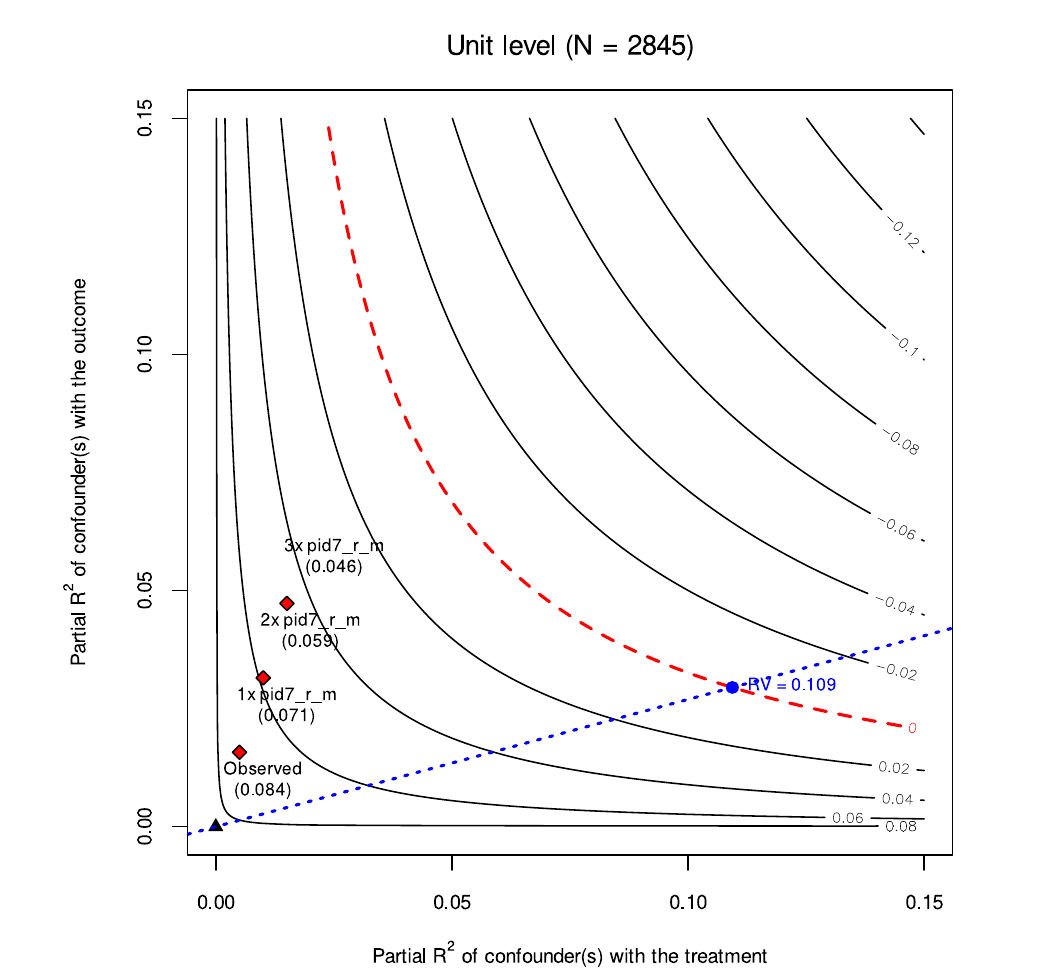}
\hfill
\includegraphics[width=0.49\linewidth]{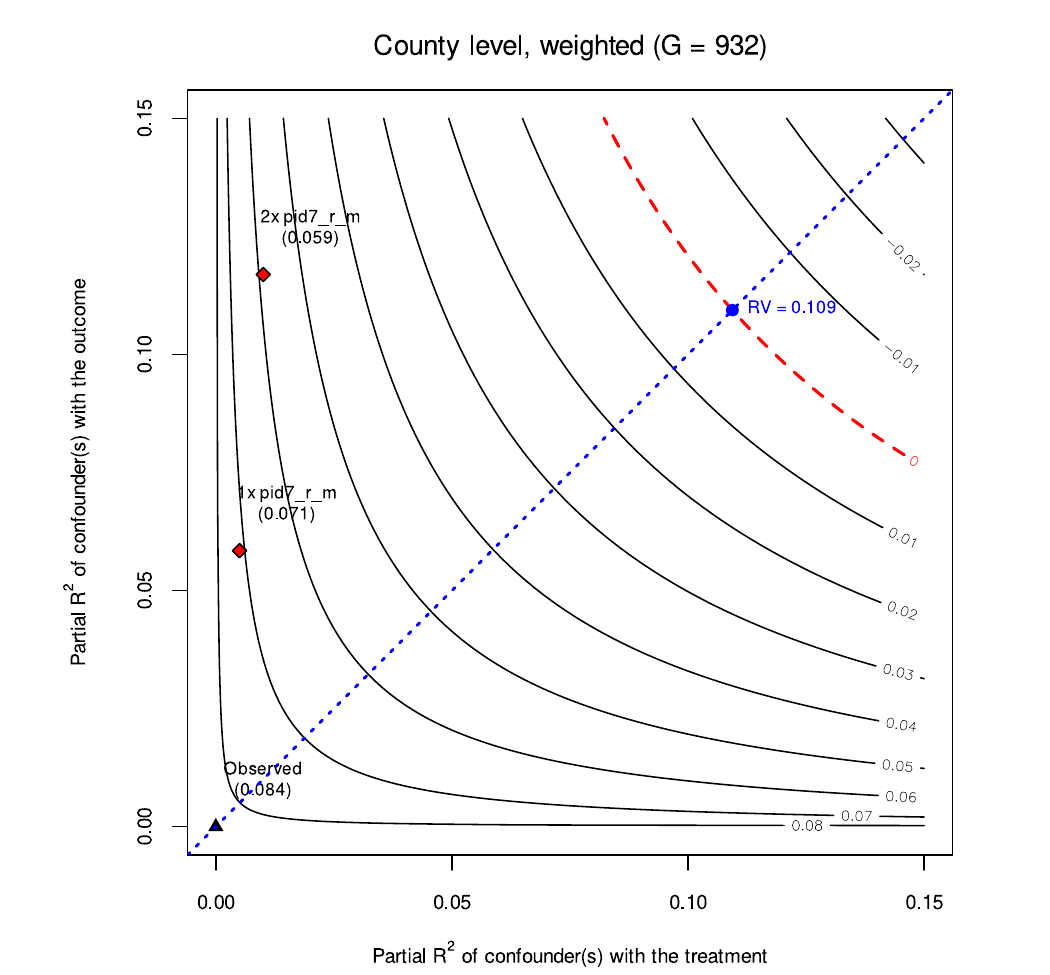}
\caption{Sensitivity contour plots for the regressions with cluster-level
covariates only}
\label{fig:contourpair}
{\par\small\emph{Note:} Dashed blue line denotes the equiconfounding line for a
cluster-level confounder. It intersects the red zero-effect contour at the
adjusted robustness value for $q=1$. Left panel denotes the regression run at the
unit level. Right panel presents the regression run at the cluster level weighted
by cluster size.}
\end{figure}

Table~\ref{tab:aggcounty} shows that the equivalent county-level regression yields
point estimates, standard errors (HC2 robust), robustness values and covariate
benchmarks identical to those from the unit-level regression.
Figure~\ref{fig:contourpair} presents the contour plots from these two analyses
side-by-side, showing that the only difference between the two is essentially a
re-scaling of the relevant Y-axis from $[0, \eta^2_{Y \mid D, \bu{X}}]$ on the
unit-regression side to $[0, 1]$ on the cluster-regression side.

The replication is consistent with the broader conclusion of this paper. Although
properly accounting for the between-cluster variation in the covariates attenuates
the point estimates slightly, correcting the reported \texttt{sensemakr}
robustness statistics to account for clustering of treatment suggests that the
results are actually more robust to omitted confounding than originally presented.
A confounder that genuinely explains an equivalent amount of outcome and treatment
variation would need to explain $10.9\%$ of that variation to eliminate the
effect. The $7.4\%$ figure from \citet{reny2026rising} was conservative in that it
incorrectly attributed explanatory power to the confounder that it could never
possibly have.

\section{Conclusion}
\label{sec:conclusion}

Analyzing the effects of cluster-level treatments using individual-level
regressions remains common in applied work. The results of this paper show that
applying a \emph{unit-level} sensitivity analysis for omitted variable bias leads
to much more \emph{conservative} assessment of robustness when treatment
assignment is clustered. Standard practice is, if anything, overly pessimistic.
This suggests some cause for relief as applying a unit-level analysis in a setting
where treatment was actually clustered will not make researchers overconfident in
the robustness of their findings. As \citet{hansen2014clustered} note, there are
many cases where such an analytical decision is necessary. For example, clusters
may not be observed directly due to deidentification in public-use datasets. While
this also poses a separate challenge for statistical inference, the results for
sensitivity to unobserved confounding are more promising. The findings here echo
the earlier conclusions of \citet{hansen2014clustered} that clustered designs are
less sensitive to unobserved confounding as a hypothetical confounder is
restricted in how it can act on the observations. Confounding can only explain
variation between clusters, not within.

These results also have straightforward implications for extensions of the
\citet{cinelli2020making} OVB formula to the instrumental variables setting
\citep{cinelli2025omitted} and to the generalization of the linear model OVB to
the partially linear model (and to other nonparametric causal models) via the
Riesz representation \citep{chernozhukov2022long}. The ratio of the
between-cluster variance to the total variance serves as a further bound on how
much bias a hypothetical cluster-level outcome confounder could produce.

An interesting and perhaps more ubiquitous case of applying a unit-level
sensitivity analysis to a clustered-treatment design is the two-way fixed effects
(TWFE) regression estimator for differences-in-differences (DiD). It remains
common in applied work to estimate causal effects for DiD designs using a
unit-time TWFE linear model with group and time fixed effects parameters. Because
the estimator is a typical OLS regression, researchers often apply the
\citet{cinelli2020making} regression OVB to the TWFE model as a way of assessing
robustness to a hypothetical confounder that induces a violation of parallel
trends.\footnote{Only recently has the OVB-style sensitivity analysis been
formally applied to the difference-in-differences setting. See
\citet{wang2026omitted} which extends the approach of
\citet{chernozhukov2022long} to the doubly-robust difference-in-differences
estimator \citep{sant2020doubly, callaway2021difference}.} With a single treatment
initiation time, TWFE is identical to a simple $2 \times 2$
difference-in-differences and has an equivalent \emph{cluster-level} regression
which regresses the first-difference in the average outcome observed before and
after treatment on an indicator for treatment. There is a similar discrepancy in
applying regression OVB sensitivity to the TWFE and first-differences regressions,
with the latter avoiding the deflation factors identified here. The cluster-level
regression averages over \emph{two} elements of the unit-level regression that
exhibit no variation in treatment: repeated observations over time in the
pre-treatment and post-treatment windows and clustering in treatment assignment
(as in a repeated cross-section design).\footnote{Beyond the problem of
clustering, first-differences regressions also allow researchers to more easily
include and benchmark against time-invariant threats to parallel trends. These are
commonly omitted from TWFE regressions due to collinearity with the unit fixed
effects \citep{caetano2024difference}.} Researchers should therefore apply similar
caution when interpreting results the \citet{cinelli2020making} sensitivity
analysis for TWFE regressions.

I have focused here only on sensitivity for the point estimate, although
\citet{cinelli2020making} also consider sensitivity for upper and lower confidence
bounds to evaluate the minimum amount of confounding needed to make a result
``insignificant'' rather than ``zero''. However, the expressions derived for this
assume a \emph{homoskedastic} error structure and use the conventional OLS
estimate of the standard error using the pooled standard deviation of the
residuals. In practice, most modern regression analyses use a
heteroskedasticity-robust variance estimator due to the implausibility of the
constant error variance assumption. In the context of clustered treatment
assignment, valid inference from a unit-level regression requires a cluster-robust
variance estimator that permits error correlation among units that share a cluster
\citep{abadie2023should}. Currently, there is no such extension of the OVB
sensitivity analysis to either form of robust confidence intervals.

This paper has proposed simple corrections to regression OVB sensitivity analysis
when analyzing clustered designs with unit-level regressions, but perhaps the
simplest solution for applied researchers is to just run the regression at the
level of treatment assignment. There appear to be minimal drawbacks from doing so
and, as shown in this paper, a risk of over-optimism when benchmarking against
unit-level covariates included without their covariate means. While there are
efficiency drawbacks from an \emph{unweighted} cluster-level regression with
imbalanced clusters, the solution is to simply weight by cluster size
\citep{wooldridge2003cluster}. Both from the standpoint of identification
\emph{and} efficiency in the linear regression framework, there is no need to
include unit-level covariates net of their cluster-level summaries. If unit-level
factors influence treatment assignment, they matter \emph{only} in the degree to
which they vary between clusters and not within.

\clearpage

\printbibliography

\end{document}